\documentstyle[mprocl]{article}

\input{psfig}
\newcommand{\lapprox}{\mbox{\raisebox{-4pt}{$\,\buildrel<\over\sim\,$}}}
\newcommand{\gapprox}{\mbox{\raisebox{-4pt}{$\,\buildrel>\over\sim\,$}}}

\begin{document}

\title{MULTILEVEL BLOCKING MONTE CARLO SIMULATIONS FOR QUANTUM DOTS}

\author{R. EGGER}

\address{Institute for Theoretical Physics, University of California,\\
 Santa Barbara,
CA 93106, USA}

\author{C.H. Mak}

\address{Department of Chemistry, University of Southern California,
\\ Los Angeles, CA 90089-0482, USA}

\maketitle\abstracts{This article provides an introduction
to the ideas behind the multilevel blocking (MLB) approach to the fermion sign
problem in path-integral Monte Carlo simulations, and
also gives a detailed discussion of MLB results for quantum dots.
MLB can turn the exponential severity of the sign problem
into an algebraic one, thereby enabling numerically exact studies of 
otherwise inaccessible systems. Low-temperature simulation results for 
up to eight strongly correlated electrons in a 
parabolic 2D quantum dot are presented.
}

\section{Introduction: The fermion sign problem}

Quantum Monte Carlo (QMC) techniques are among the most powerful
methods for the computer simulation of strongly correlated
many-fermion systems, capable of delivering numerically exact results.
This article deals with a finite-temperature QMC method, namely
path-integral Monte Carlo (PIMC), which is based on 
a discretized path-integral formulation of the 
imaginary-time many-fermion propagator.
Despite its promises, applications of QMC to many-fermion
systems have been severely 
handicapped by the infamous `sign problem'.~\cite{loh}
Exchange leads to 
non-positive-definite fermionic density matrix elements, and 
the sign cancellations arising from sampling fermion paths manifest
themselves as a small signal-to-noise ratio, $\eta\sim
\exp(-N\beta E_0)$, that vanishes exponentially
with both particle number $N$ and inverse temperature $\beta=1/k_B T$
($E_0$ is a system-dependent energy scale).
Apart from variational or approximate treatments, such as the
fixed-node approximation, a solution to the sign problem in QMC 
simulations had remained elusive.

In a recent paper,~\cite{mlb}
we developed a general scheme for tackling the sign problem
in PIMC simulations. The method has been applied 
to interacting electrons in a quantum dot~\cite{egger99} and to the 
real-time dynamics of simple few-degrees-of-freedom
systems.~\cite{rtmlb} This multilevel blocking (MLB) approach
is a systematic
implementation of a simple {\em blocking strategy}.
The theorem behind the blocking strategy asserts that by sampling
groups of paths (`blocks') at the same time, 
the sign problem can always be reduced compared to 
sampling single paths as would be done normally 
(see Sec.~\ref{block} for details). 
By suitably bunching paths together
into sufficiently small blocks,
the sign cancellations among paths within the same block 
can be accounted for without the sign problem, simply because
there is no sign problem for a sufficiently small system. 
The MLB approach is then able to turn the exponential severity
of the sign problem into an algebraic one.  In practice,
that implies that significantly larger systems can be now studied
by PIMC.  

The purpose of this article is twofold.  First, we want to
present the basic ideas underlying MLB,
omitting technical implementation issues discussed in our
original work.~\cite{mlb} In particular, the reason why
blocking paths together is helpful will be discussed in some detail.
The second aim is to demonstrate the practical usefulness
of MLB for interesting quantum many-fermion applications.  
In Sec.~\ref{sec:dot},
MLB results are presented for strongly correlated 
electrons  in a 2D quantum dot.

\section{Multilevel blocking approach}

\subsection{Blocking strategy} \label{block}

We consider a many-fermion system whose state is described by a set of
quantum numbers $\vec{r}$ denoting, e.g., the positions and spins 
of {\em all}\, particles.  These quantum numbers may correspond to
electrons living on a lattice or in continuous space.
For notational simplicity, we focus
on calculating the equilibrium expectation value of a diagonal operator
or correlation function
(this can be easily generalized), 
\begin{equation} \label{fir}
\langle A\rangle = \frac{\sum_{\vec{r}} A(\vec{r})
\rho(\vec{r},\vec{r})}{\sum_{\vec{r}} 
\rho(\vec{r},\vec{r})} \;,
\end{equation}
where $\sum_{\vec{r}}$ represents either a summation for the case of a 
discrete system or an integration for a continuous system.
In PIMC applications, imaginary time is discretized into $P$ slices of 
length $\epsilon=\beta/P$.
Inserting complete sets at each slice $m=1,\ldots,P$,  
and denoting the corresponding configuration on slice $m$ 
by $\vec{r}_m$, the diagonal elements of the reduced density matrix at
$\vec{r}=\vec{r}_P$ entering Eq.(\ref{fir}) read
\begin{equation}\label{rho}
\rho(\vec{r}_P,\vec{r}_P) = \sum_{\vec{r}_1,\ldots, \vec{r}_{P-1}}
\prod_{m=1}^P \langle \vec{r}_{m+1}| e^{-\epsilon H}
|\vec{r}_m \rangle \;.
\end{equation}
One can then construct accurate analytical approximations for the
short-time propagator.
This formulation of the problem excludes 
effective actions such as those arising from an integration over the fermions
using the Hubbard-Stratonovich trans\-for\-mation,~\cite{loh}, 
since they generally lead to long-ranged imaginary-time interactions.
The MLB approach suitable for such systems is
described elsewhere.~\cite{egger1}

In dealing with a many-fermion system, we need to sum over all
particle permutations and the best way to do this is to 
antisymmetrize the short-time propagators explicitly instead of
letting the Monte Carlo handle it.  This leads to the
appearance of fermion determinants.
Strictly speaking, the antisymmetrization has to be
done only on one time slice, but the 
intrinsic sign problem is much better behaved if one antisymmetrizes
on all time slices.  

Choosing the absolute value
of the product of the short-time propagators in Eq.(\ref{rho})
as the positive definite MC weight function $P[X]$, one then 
accumulates the sign $\Phi[X]$ associated with
every path $X=(\vec{r}_1,\ldots,\vec{r}_P)$ sampled, 
\begin{equation}
\langle A \rangle =  \frac{\sum_{X} P[X] \Phi[X] A[X]} 
{\sum_{X} P[X] \Phi[X]} \;.
\end{equation}
Assuming that there are no exclusivity problems in the 
numerator so that $A[X]$ is well-behaved, we can gauge the severity of 
the sign problem in terms of the variance of the
denominator, 
\begin{equation} \label{var}
\sigma^2 \approx \frac{1}{N_s} \left( \langle \Phi^2 \rangle
- \langle \Phi \rangle^2 \right)   \;,
\end{equation}
where $N_s$ is the number of MC samples taken and the stochastic averages are 
calculated with $P$ as the weight function. 
For the fermion sign problem, where $\Phi=\pm 1$ and hence
$\langle \Phi^2\rangle = 1$,  the variance of the 
signal is controlled by the size of $|\langle\Phi\rangle|$.

Remarkably, one can achieve considerable progress by 
simply {\em blocking paths together}.~\cite{makacp} By this we mean 
instead of sampling single paths in the MC, we
can sample sets of paths (``blocks'').
Under such a blocking operation,
the stochastic estimate for $\langle A \rangle$
takes the form
\begin{equation} \label{start2}
\langle A \rangle =  \frac{\sum_{B} \left(
\sum_{X\in B} P[X] \Phi[X] A[X] \right)} 
{\sum_{B} \left(\sum_{X\in B} P[X] \Phi[X] \right)} \;,
\end{equation}
where one first sums over the configurations belonging to a 
block $B$ in a way that is not affected by the sign problem, and then
stochastically sums over the blocks.  The summation within a block
must therefore be done non-stochastically, or alternatively the
block size must be chosen sufficiently small.
Of course, there is considerable freedom in how to choose this
blocking.

Let us analyze the variance $\sigma^{\prime 2}$
of the denominator of Eq.(\ref{start2}). We first define new 
sampling functions 
in terms of the blocks which are then sampled stochastically,
\begin{equation}
 P'[B] = \left| \sum_{X \in B} P[X] \Phi[X] \right| \;,\quad
\Phi'[B] ={\rm sgn} \left(\sum_{X \in B} P[X] \Phi[X]\right) \;.
\end{equation}
Rewriting the average sign in the new representation, i.e.,
using $P'[B]$ as the weight, then inserting
the definition of $P'$ and $\Phi'$ in the numerator,
\[
\langle \Phi'[B] \rangle = 
\frac{\sum_{B} P'[B] \Phi'[B]}{ \sum_B P'[B]}
= \frac{\sum_{X} P[X] \Phi [X]}{ \sum_B P'[B]} \;,
\]
and comparing to the average sign in the standard 
representation using $P[X]$ as the weight, we obtain
\[
\frac{|\langle \Phi' \rangle|}{ |\langle \Phi \rangle|} =
\frac{\sum_{X} P[X]} {\sum_{B} P'[B]} \;.
\]
By virtue of the Schwarz inequality,  
\[
\sum_{B} P'[B] = \sum_{B} \left| \sum_{X\in B} 
P[X] \Phi[X] \right|\leq \sum_{B} \left| \sum_{X\in 
B} P[X] \right|  = \sum_{X} P[X]\;,
\]
we see that {\em for any kind of blocking}, the average sign
improves (or stays the same),
$|\langle\Phi'\rangle|
\geq | \langle \Phi \rangle|$.
Furthermore, since $\langle \Phi^{\prime 2} \rangle
= \langle \Phi^{2} \rangle = 1$, we conclude from Eq.(\ref{var}) that 
\begin{equation} \label{eqi} 
\sigma^{\prime\,2} \leq \sigma^2\;,
\end{equation} 
and hence {\em the signal-to-noise ratio is always improved upon 
blocking configurations together}.  Clearly,
the worst blocking one could possibly 
choose would be to group the configurations into two blocks, one with
positive sign and the other with negative sign. In this 
case, blocking yields no improvement whatsoever, and the `$\leq$'
becomes `$=$' in Eq.(\ref{eqi}).  It is apparent from Eq.(\ref{eqi}) 
that the blocking strategy provides a systematic approach to 
reduce the sign problem.   


\subsection{Multilevel blocking approach}

A direct implementation of the strategy described in Sec.~\ref{block} 
does indeed improve the sign problem but will not remove its exponential
severity.  The reason is simply that for a sufficiently large
system, there will be too many blocks, and once the 
signals coming from these blocks are allowed to interfere,
one again runs into the sign problem (albeit with a smaller
scale $E_0$).   The resolution to this problem comes from
the multilevel blocking (MLB) approach~\cite{mlb} 
where one {\em applies the
blocking strategy in a recursive manner to the blocks} again.
In a sense, we form new blocks containing a sufficiently 
small number of elementary ones, and repeat this process 
until only one block is left.

Instead of allowing for an uncontrolled
interference of the block signals, we therefore 
subdivide the block space by forming a hierarchy of
`levels' $\ell=0,\ldots, L$, where the Trotter number
must be of the form $P=2^L$. 
All elementary blocks are distributed onto these levels,
and a MC sweep proceeds from the bottom level $\ell=0$ up to
the top level $\ell=L$.  When sampling on a given level $\ell<L$,
no sign problem is present if sufficiently small block sizes
have been chosen. The nontrivial computational task
consists of finding a controllable way of transferring 
interference information from level $\ell$ to $\ell+1$.
 It is then indeed possible to proceed without
numerical instabilities from the bottom up to the top level,
where the expectation values of interest are computed.

The blocks at the top level still have different signs, 
but the interference is however much weaker than in the original
formulation.  Empirically, we found that the exponential
severity is turned into an only algebraic one. The
algebraic scaling can also be derived by a simple
argument.~\cite{egger1}
For lack of space, we refer  the interested reader to 
our original paper~\cite{mlb} for practical implementation
issues of the MLB algorithm, and now turn to an application.

\section{Application: Quantum dots}\label{sec:dot}

Quantum dots are solid-state artificial atoms with tunable properties.
Confining a small number of electrons $N$ in a 2D
electron gas in semiconductor heterostructures,
novel effects due to the interplay
between confinement and the Coulomb interaction  have been 
observed experimentally.~\cite{ashoori}
For small $N$, comparison of experiments to the generalized Kohn
theorem indicates that the confinement potential is parabolic
and hence quite shallow compared to conventional atoms.
Employing the standard electron gas parameter $r_s$ to
quantify the correlation strength,
a Fermi-liquid-like picture with the sequential filling of single-particle 
orbitals is applicable only for small $r_s$.  
In the low-density (strong-interaction)
limit of large $r_s$, however, classical considerations
suggest a Wigner crystal-like phase
with electrons spatially arranged in shells,~\cite{classical}
termed {\em Wigner molecule} due to its finite extent.
We are particularly interested in the crossover regime between
these two limits, where single-particle or classical descriptions 
break down, and basically no other sufficiently accurate
method is available. 
Exact diagonalization is limited to very small $N$
since one otherwise introduces a
huge error due to the truncation of the Hilbert space.
Hartree-Fock (and related) calculations become unreliable
for large $r_s$ and incorrectly favor spin-polarized states.
Furthermore, density functional calculations can introduce
uncontrolled approximations.  In fact, by comparing with 
our numerically exact data,  
symmetry-broken spin density solutions
found in a recent density functional study,~\cite{reimann}
and later in a Hartree-Fock calculation,~\cite{landman} 
are most likely artefacts of the approximations involved.
A variational Monte Carlo with a fixed-node approximation 
has been employed by Bolton.~\cite{bolton}   Comparing this
to our exact calculations, we find that typical fixed-node
errors in the total energy for $N>5$ are of the order of 10$\%$. 
It is then clear that one must resort to exact methods,
especially when examining spin-dependent quantities.

\subsection{Model}

We study  a clean 2D parabolic quantum dot with zero magnetic field,
\begin{equation}
H= \sum_{j=1}^N \left(\frac{\vec{p}_j{}^2}{2m^*} +
\frac{m^*\omega_0^2}{2} \vec{x}_j{}^2 \right) +
\sum_{i<j=1}^N \frac{e^2}{\kappa|\vec{x}_i -\vec{x}_j|} \;.
\end{equation}
Here the electron positions (momenta) are
given by $\vec{x}_j \; (\vec{p}_j)$, their effective mass is $m^*$,
and the dielectric constant is $\kappa$. 
The MLB calculations are carried out at fixed $N$
and fixed $z$-component of the total spin, 
$S=(N_\uparrow- N_\downarrow)/2$.
We have studied the energy,
$E = \langle H \rangle$, the radial charge and spin
densities $\rho(r)$ and $s_z(r)$ normalized to $\int_0^\infty dr \,
2\pi r \, \rho(r) = N$ and $\int dr \, 2\pi r \,
s_z(r) = S$, and the two-particle correlation function
\begin{equation} \label{cr}
g_S^{}(\vec{x}) = \frac{2\pi l_0^2}{N(N-1)}
\left \langle \sum_{i\neq j=1}^N \delta(\vec{x}-\vec{x}_i+
\vec{x}_j )\right \rangle \;.
\end{equation}
$g_S^{}$ is isotropic, and with 
$y=r/l_0$ prefactors are chosen such that $\int_0^\infty dy \,
y g^{}_S(y)=1$.  The confinement length scale $l_0=\sqrt{\hbar/m^*\omega_0}$
allows the interaction strength to be parametrized by $\lambda=l_0/a
=e^2/\kappa\omega_0 l_0$, where $a$
is the effective Bohr radius of the artificial atom.
For any given $N$ and $\lambda$,
the parameter $r_s=r^*/a$
follows by identifying $r^*$ with
 the first maximum in $\sum_S g^{}_S(r)$.
In the simulations, unless noted otherwise,
the temperature was set to $k_B T = 0.1\, \hbar
\omega_0$.
Data were collected from several $10^4$ samples
for each parameter set $\{N,S,\lambda\}$, with a
typical CPU time requirement of
a few days (for each set) on a SGI Octane workstation.
To check our code, we have accurately 
reproduced the known exact solution for $N=2$.

\subsection{Charge and spin densities}

\begin{figure}
\psfig{figure=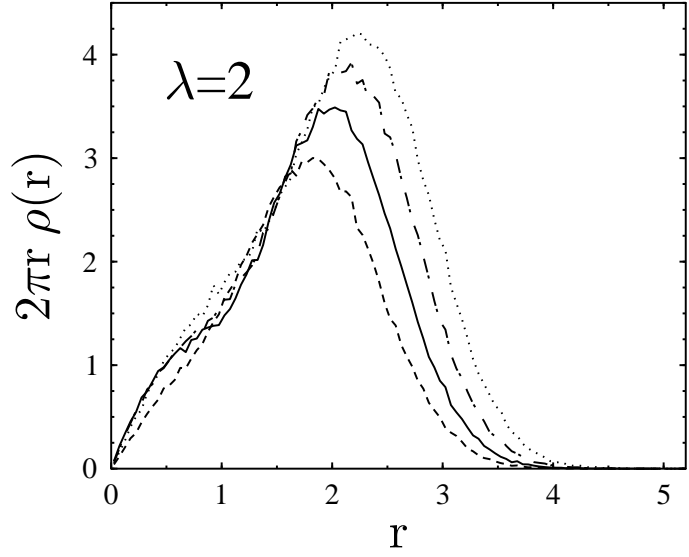,height=3.5in}
\psfig{figure=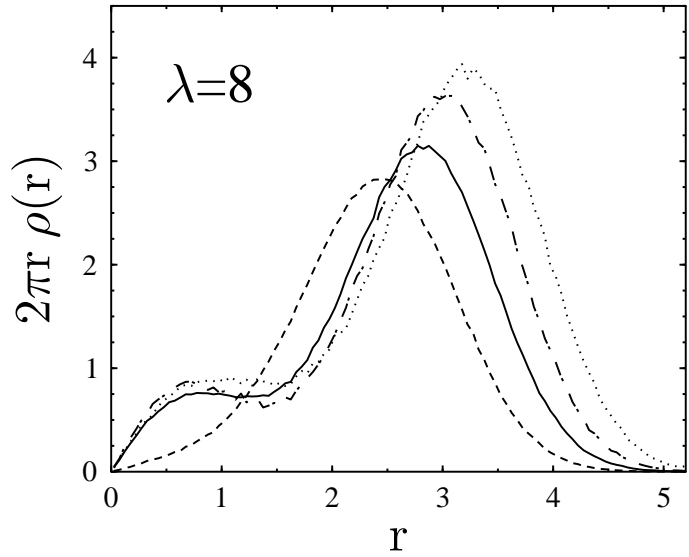,height=3.5in}
\caption[]{\label{fig1}
Density $\rho(r)$ of the spin-polarized state ($S=N/2$)
for $\lambda=2$ and $\lambda=8$.
Dashed, solid, dashed-dotted, and dotted curves correspond to $N=5$,
6, 7, and 8, respectively. Units are such that $l_0=1$.
}
\end{figure}

Figure \ref{fig1} shows the radial charge density $\rho(r)$ of the
spin-polarized state $S=N/2$ for $N=5$ to 8 electrons. 
For $\lambda=2$,
increasing $N$ does not change $\rho(r)$ qualitatively, but
the situation is different for strong interactions ($\lambda=8$),
where one can observe shell formation in real space. Such a
spatial structure is clear evidence for Wigner molecule behavior.
The classical shell filling sequence has
been computed recently.~\cite{classical}
For $N<6$, the electrons arrange on a ring, but the sixth electron
then goes into the center (shell filling 1-5). 
Furthermore, electrons 7 and 8 enter the outer ring again.
These predictions are in accordance with our data. 
Clear indications of a spatial shell structure at $N\geq 6$
can be observed even for
$\lambda = 4$, albeit quantum fluctuations tend to wash them
out.
For $\lambda \gapprox 4$, the charge densities are 
basically insensitive to $S$.  
This is characteristic for a classical Wigner crystal, where
the Pauli principle and spin-dependent properties are
of secondary importance.
 Our numerical results for the spin density
in this regime simply follow the corresponding charge density
according to $s_z(r)\simeq (S/N) \rho(r)$. 
A significant $S$-dependence of charge and
spin densities is observed only for weak correlations.

\subsection{Crossover from Fermi liquid to Wigner molecule}

To study the crossover from weak to strong correlations, we employ
the `spin sensitivity', normalized to unity for $r_s=0$,
\begin{equation} \label{xi}
\xi_N(r_s) \propto
 \sum_{S,S'} \int_0^\infty dy \; y \, | g^{}_S(y)-g^{}_{S'}(y) | \;.
\end{equation}
The correlation function $g^{}_S(r)$ in Eq.(\ref{cr})
is a very sensitive measure of Fermi statistics,
in particular revealing the spin-dependent correlation hole.  
As interactions tend to wash out the Fermi surface,
the quantity (\ref{xi}) is largest for a Fermi gas, $r_s=0$.
Since for $r_s\to \infty$,  
$g_S(r)$ becomes completely spin-independent,
$\xi_N(r_s)$ decays from unity at $r_s=0$ down to
zero as $r_s\to \infty$.  The functional dependence of this decay
provides insight about the crossover phenomenon under study.

\begin{figure}
\hfil
\psfig{figure=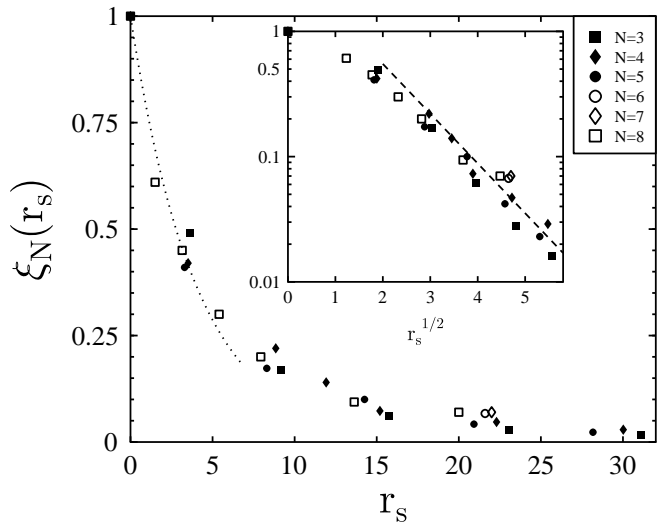,height=3.2in}
\hfil
\caption[]{\label{fig2} Numerical results for $\xi_N(r_s)$.
Statistical errors are of the order of the symbol size. The
dotted curve, given by $\exp(-r_s/r_c)$ with $r_c=4$, is a guide
to the eye only. The inset shows the same data on a
semi-logarithmic scale as a function of $\sqrt{r_s}$. The
dashed line is given by Eq.~(\ref{sc}).}
\end{figure}

Figure \ref{fig2} reveals that the function 
$\xi_N(r_s)$ becomes remarkably
{\sl universal}\, and depends only very weakly on $N$. Its decay
defines a crossover scale $r_c$, where an exponential
fit for small $r_s$ yields $r_c\approx 4$.
For $r_s>4$, the data can be fitted by
\begin{equation}\label{sc}
\xi(r_s) \sim \exp\left(-\sqrt{r_s/r_c^\prime}\right) \;,
\end{equation}
where $r_c^\prime\approx 1.2$.   Remarkably, this is
precisely the behavior expected from a semiclassical WKB estimate
for a Wigner molecule.~\cite{egger99}
The crossover value $r_c\approx 4$ is also consistent with the onset of
spatial shell structures in the density, and with 
the spin-dependent ground state energies expected for a Wigner molecule.
Therefore the crossover from weak to strong correlations
is characterized by the surprisingly small value $r_c\approx 4$, instead of
$r_c\approx 37$ found for the bulk 2D electron gas.~\cite{tanatar}
The stabilization of the Wigner molecule can be 
ascribed to the confinement potential. In the 
thermodynamic limit,  $\omega_0\to 0$ with $r_s$ fixed, 
plasmons govern the low-energy physics, and hence
the bulk value $r_c\approx 37$ becomes relevant for very large $N$. 
For GaAs based quantum dots, we estimate that for $N\lapprox 10^4$,
the value $r_c\approx 4$ is valid.
Remarkably, very recent experiments on vertical quantum dots~\cite{ash3}
have found evidence for an even smaller crossover scale $r_c\approx 1.8$.  
The experimental study was carried out in a magnetic
field, and the dot contained several impurities.  Since
both effects tend to stabilize a Wigner crystallized phase,
our theoretical prediction and the experimental observation
appear to be consistent with each other, especially since
we are concerned with a rather smooth crossover phenomenon.

\subsection{Spin-dependent energies}

\begin{table}[t]
\caption{\label{table1}
MLB data for the energy for various $\{ N, S, \lambda\}$
parameter sets. Bracketed numbers denote statistical errors.
}
\vspace{0.2cm}
\begin{center}
\footnotesize
\begin{tabular}{|llll||llll|} \hline
$N$ & $S$ & $\lambda$ & $E/\hbar\omega_0$&$N$ & $S$ & $\lambda$ &
$E/\hbar\omega_0$\\ \hline
3 & 3/2 & 2 & 8.37(1) & 5 & 5/2 & 8 & 42.86(4) \\
3 & 1/2 & 2 & 8.16(3) &  5 & 3/2 & 8 & 42.82(2) \\
3 & 3/2 & 4 & 11.05(1) & 5 & 1/2 & 8 & 42.77(4) \\
3 & 1/2 & 4 & 11.05(2) & 5 & 5/2 & 10 & 48.79(2) \\
3 & 3/2 & 6 & 13.43(1) & 5 & 3/2 & 10 & 48.78(3) \\
3 & 3/2 & 8 & 15.59(1) & 5 & 1/2 & 10 & 48.76(2) \\
3 & 3/2 & 10 & 17.60(1) & 6 & 3 & 8 & 60.42(2) \\
4 & 2 & 2 & 14.30(5) & 6 & 1 & 8 & 60.37(2) \\
4 & 1 & 2 & 13.78(6) & 7 & 7/2 & 8 & 80.59(4) \\
4 & 2 & 4 & 19.42(1) & 7 & 5/2 & 8 & 80.45(4) \\
4 & 1 & 4 & 19.15(4) &  8 & 4 & 2 & 48.3(2) \\
4 & 2 & 6 & 23.790(12) & 8 & 3 & 2 & 47.4(3) \\
4 & 1 & 6 & 23.62(2) & 8 & 2 & 2 & 46.9(3) \\
4 & 2 & 8 & 27.823(11) & 8 & 1 & 2 & 46.5(2) \\
4 & 1 & 8 & 27.72(1) & 8 & 4 & 4 & 69.2(1) \\
4 & 2 & 10 & 31.538(12) & 8 & 3 & 4 & 68.5(2) \\
4 & 1 & 10 & 31.48(2) & 8 & 2 & 4 & 68.3(2) \\
5 & 5/2 & 2 & 21.29(6) & 8 & 4 & 6 & 86.92(6) \\
5 & 3/2 & 2 & 20.71(8) &  8 & 3 & 6 & 86.82(5) \\
5 & 1/2 & 2 & 20.30(8) & 8 & 2 & 6 & 86.74(4) \\
5 & 5/2 & 4 & 29.22(7) & 8 & 4  & 8  & 103.26(5)  \\
5 & 3/2 & 4 & 29.15(6) & 8 & 3  & 8  & 103.19(4)  \\
5 & 1/2 & 4 & 29.09(6) & 8 & 2 & 8 & 103.08(4)\\
5 & 5/2 & 6 & 36.44(3) &  &  &  &  \\
5 & 3/2 & 6 & 36.35(4) &  &  &  & \\
5 & 1/2 & 6 & 36.26(4) &  &  &  & \\ \hline
\end{tabular}
\end{center}
\end{table}

 MLB results for the
energy at different parameter sets $\{N,S,\lambda\}$ are
listed  in Table \ref{table1}.  For given $N$ and $\lambda$, if the ground
state is (partially) spin-polarized with spin $S$, 
the simulations should yield the same energies for all $S'<S$. 
Within the accuracy of the calculation, this consistency check
is indeed fulfilled.
For strong correlations, $r_s>r_c$, the spin-dependent energy
levels differ substantially from a
 single-particle orbital picture.
In particular, the ground-state spin $S$ can change and
the relative energy of higher-spin states becomes much
smaller than $\hbar\omega_0$.

\begin{figure}
\hfil
\psfig{figure=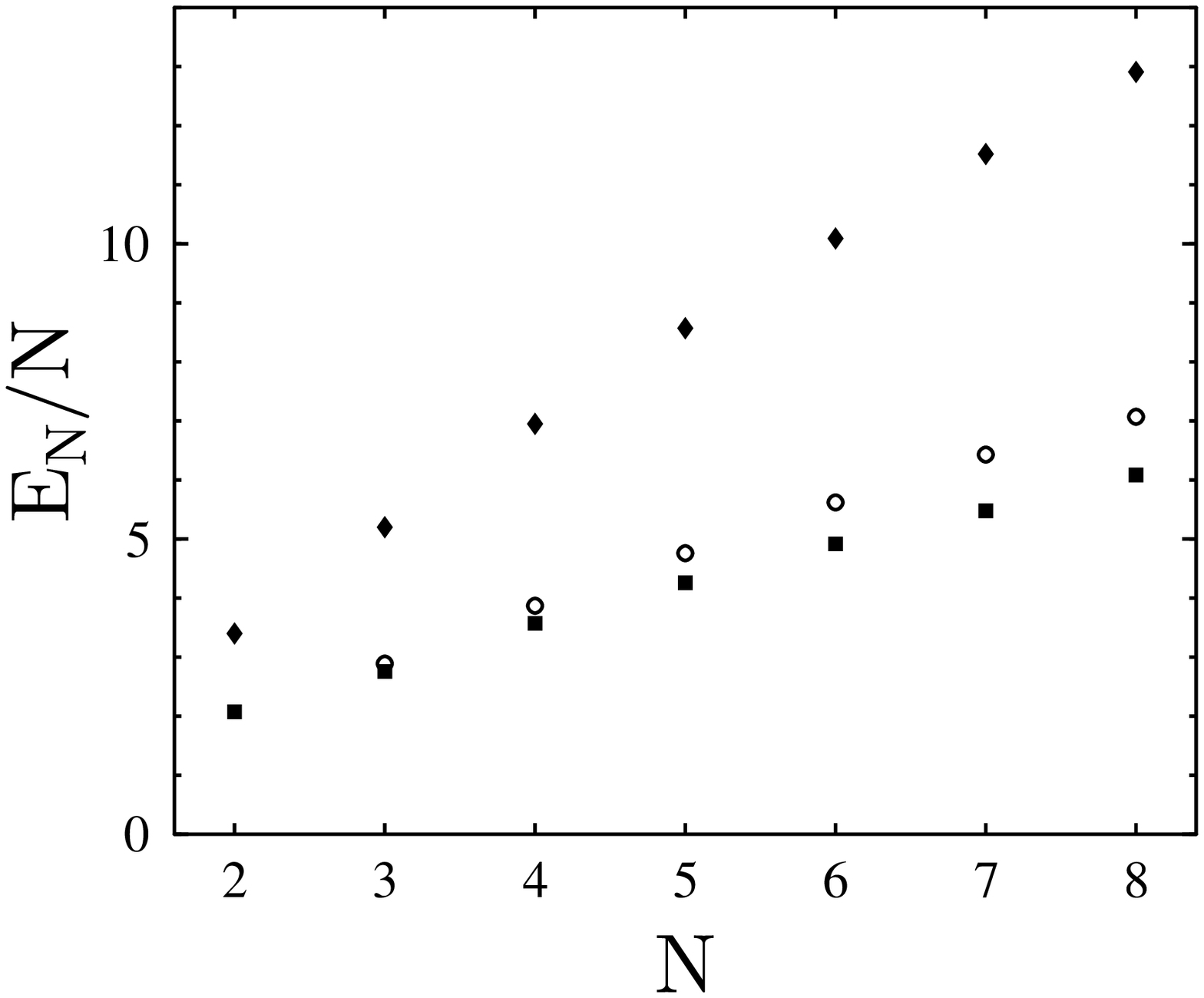,height=3.2in}
\hfil
\caption[]{\label{fig3}
Energy per electron, $E_N/N$, for $S=N/2$ and $\hbar\beta\omega_0=6$,
in units
of $\hbar \omega_0$, for $\lambda=2$ (squares) and $\lambda=8$ (diamonds).
Statistical errors are smaller than the symbol size.
Open circles are $T=0$ fixed-node QMC results~\cite{bolton} for 
$\lambda=2$.}
\end{figure}

For $N=3$ electrons, as $r_s$ is increased, a transition occurs from
$S=1/2$ to  $S=3/2$ at an interaction strength
$\lambda\approx 5$ corresponding to $r_s\approx 8$.  
For $N=4$, we encounter a Hund's rule case with a small-$r_s$ ground state
characterized by $S=1$.
>From our data, this standard Hund's rule 
covers the full range of $r_s$.
A similar situation arises for $N=5$, where 
the ground state is characterized by $S=1/2$ for all $r_s$.
Turning to $N=6$, while one has filled orbitals and
hence a zero-spin ground state for weak correlations,
for $\lambda=8$ we find
a $S=1$ ground state.  A similar transition
from a $S=1/2$ state for weak correlations
to a partially spin-polarized $S=5/2$ state is 
found for $N=7$. Finally, for $N=8$, 
as expected from Hund's rule,
a $S=1$ ground state is observed for small $r_s$. 
 However, for $\lambda \gapprox 4$, corresponding
to $r_s\gapprox 10$, the ground state spin changes to $S=2$,
implying a different `strong-coupling' Hund's rule.

Let us finally address the issue of `magic numbers'. For small $r_s$,
the filling of orbital shells and Hund's rule arguments predict 
that certain $N$ are exceptionally stable. 
Results for the energy per electron, $E_N/N$, in the spin-polarized state
$S=N/2$, are shown in Figure \ref{fig3}.
Notably, there are no obvious cusps or breaks in the
$N$-dependence of the energy. 
Our $\lambda=2$ data in Fig.~\ref{fig3} suggest that an 
explanation of the experimentally
observed magic numbers~\cite{tarucha} has to involve spin and/or 
magnetic field effects.  Remarkably, the absence of 
pronounced cusps in $E_N/N$ for strong correlations ($\lambda=8$) 
is in accordance with the classical analysis.~\cite{classical}
Therefore magic numbers seem to play only a minor role in
the Wigner molecule phase.

\vspace*{-2pt}

\section*{Acknowledgments}
We acknowledge collaborations and discussions 
with Ray Ashoori, Hermann Grabert, Wolfgang H\"ausler, Boris Reusch, and
Viktor Sverdlov.
This research has been supported by the SFB 276 of the Deutsche
Forschungsgemeinschaft (Bonn), 
by the National Science Foundation
under grants CHE-9528121, CHE-9970766 and PHY94-07194, 
by the Sloan Foundation,
and by the Dreyfus Foundation.

\eject


\vspace*{-9pt}

\section*{References}
\begin{thebibliography}{99}
\bibitem{loh} E.Y. Loh Jr., J. Gubernatis, R.T. Scalettar,
S.R. White, D.J. Scalapino, and R.L. Sugar,  {\em Phys. Rev. B}
{\bf 41}, 9301 (1990). 
\bibitem{mlb} C.H.~Mak, R.~Egger, and H.~Weber-Gottschick,
{\em Phys. Rev. Lett.} {\bf 81}, 4533 (1998).
\bibitem{egger99} R. Egger, W. H\"ausler, C.H. Mak, and H. Grabert,
{\em Phys. Rev. Lett.} {\bf 82}, 3320 (1999); {\bf 83}, 462(E) (1999).
\bibitem{rtmlb} C.H.~Mak and R.~Egger, 
{\em J. Chem. Phys.} {\bf 110}, 12 (1999). 
\bibitem{egger1} R. Egger and C.H. Mak, {\em Phys. Rev. E} (submitted).
\bibitem{makacp} C.H. Mak and R. Egger, in: New Methods in Computational
Quantum Mechanics, {\em Adv. Chem. Phys.} {\bf 93}, 39 (1996).
\bibitem{ashoori} R.C. Ashoori, {\em Nature} {\bf 379}, 413 (1996).
\bibitem{classical}
F. Bolton and U. R\"ossler, {\em Superlatt. Microstruct.} {\bf 13}, 139 (1993);
V.M. Bedanov and F.M. Peeters, {\em Phys. Rev. B} {\bf 49}, 2667 (1994).
\bibitem{reimann} M. Koskinen {\em et al.}, {\em Phys. Rev. Lett.}
{\bf 79}, 1389 (1997).
\bibitem{landman} C. Yannouleas and U. Landman, {\em
Phys. Rev. Lett.} {\bf 82}, 5325 (1999). 
\bibitem{bolton} F. Bolton, {\em Phys. Rev. Lett.} {\bf 73}, 158 (1994).
\bibitem{tanatar} B. Tanatar and D.M. Ceperley, {\em Phys. Rev. B} {\bf 39},
5005 (1989).
\bibitem{ash3} N.B. Zhitenev {\em et al.}, 
{\em Science} {\bf 285}, 715 (1999). 
\bibitem{tarucha} S. Tarucha {\em et al.}, {\em Phys. Rev. Lett.}
{\bf 77}, 3613 (1996).
\end{thebibliography}
\end{document}